# Planar Hall effect in type-II Weyl semimetal WTe$_2$

Y. J. Wang,[1,2] J. X. Gong,[1] D. D. Liang,[1] M. Ge,[2] J. R. Wang,[1] W. K. Zhu,[1,3,*] and C. J. Zhang,[1,3,4,†]

[1]*Anhui Province Key Laboratory of Condensed Matter Physics at Extreme Conditions, High Magnetic Field Laboratory, Chinese Academy of Sciences, Hefei 230031, China*

[2]*Hefei National Laboratory for Physical Sciences at Microscale, University of Science and Technology of China, Hefei 230026, China*

[3] *Institute of Physical Science and Information Technology, Anhui University, Hefei 230601, China*

[4] *Collaborative Innovation Center of Advanced Microstructures, Nanjing University, Nanjing 210093, China*

[*]wkzhu@hmfl.ac.cn

[†]zhangcj@hmfl.ac.cn

Adler-Bell-Jackiw chiral anomaly is a representative feature arising from the topological nature in topological semimetal. We report the first experimental observation of giant planar Hall effect in type-II Weyl semimetal WTe$_2$. Our comprehensive analyes of the experimental data demonstrate that the detected planar Hall effect is originated from the chiral anomaly of Weyl fermions. Unlike the somewhat elusive negative magnetoresistance, the planar Hall effect is robust and easy to be detected in type-II Weyl semimetal. This work reveals that the planar Hall effect is an effective transport probe to determine the topological nature of topological semimetals, especially in type-II Weyl semimetals.

**I. INTRODUCTION**

Topological semimetals (TSMs) represent quantum materials that host crossing and linearly dispersive energy bands in momentum space [1-5], and attract accumulating interest due to their physical significance and potential technological applications. In Dirac semimetals, the band crossings (i.e., Dirac nodes) are protected by both inversion symmetry and time reversal symmetry [3-5]. When inversion or time reversal symmetry is broken, the Dirac points could be splitted into pairs of Weyl points and the Dirac semimetal becomes a Weyl semimetal [2, 3]. In recent years, a number of compounds have been theoretically predicted and experimentally demonstrated to be topological semimetals[4-12].

However, the experimental tools to probe and determine the topological semimetal states are always found very limited. When Dirac/Weyl points are close to the Fermi level, electrons behave like Dirac/Weyl fermions and some unusual phenomena emerge, such as small effective carrier mass, high carrier mobility, nontrivial Berry phase [13, 14], Fermi arcs[1, 7-10] and Adler-Bell-Jackiw chiral anomaly [15, 16]. Usually, cone-like structures (or Fermi arcs) observed in angle-resolved photoemission spectroscopy (ARPES), and chiral anomaly detected in transport measurements are solid evidence for the determination of TSM states. Chiral anomaly, originating from the non-conservation of particle numbers under parallel electric and magnetic fields in TSMs [15-17], can induce negative magnetoresistance (NMR) [15, 16, 18], anomalous Hall effect [19, 20], and nonlocal transport [21]. The chiral anomaly induced NMR has been claimed as a valid approach to confirm the topological character in a number of TSMs, including $Cd_3As_2$ [17, 22], $Na_3Bi$ [23], $ZrTe_5$ [24], TaAs [14], $WTe_2$ [25], GdPtBi [26], etc.

Generally speaking, TSMs have very large normal magnetoresistance, so the very small NMR may be concealed by a slightly imperfect misalignment of the magnetic field and the current in the sample, which makes the NMR confined within a narrow angular window. Moreover, the chiral anomaly induced conductance is inversely proportional to the square of chemical potential measured from the energy of Dirac/Weyl points. If the Dirac/Weyl points are far away from Fermi level, the chiral anomaly induced NMR will be too weak to be observed. Such two factors increase the difficulty in detecting the chiral anomaly induced NMR in TSMs, especially in type-II Weyl semimetals (e.g., $WTe_2$ and $MoTe_2$), where the NMR can only be observed along specific directions [12, 25] and in samples of special component [27]. Besides, the negative MR effect can also be induced by other mechanisms, such as current jetting [26], suppression of magnetic scattering [28], weak localization effect [29] and so on, which will reduce the validity of NMR measurements, so new tools to confirm the chiral anomaly in TSMs are required.

Recently, it is proposed that a giant planar Hall effect (PHE) could occur in TSMs [30, 31]. The PHE is induced by the chiral anomaly and nontrivial Berry curvature, thus the PHE could be regarded as a solid transport signature of topological quantum states. As a matter of fact, the giant PHE phenomena have been experimentally revealed in 3D Dirac semimetal $Cd_3As_2$ and type-I Weyl semimetal GdPtBi [32-34]. Also note that the PHE can be realized in ferromagnetic metals [35], which arises from the interplay of magnetic order and spin-orbit interactions, but its value is very small.

In this paper, we report the observation of PHE in type-II Weyl semimetal $WTe_2$ [11, 12, 25, 36]. Considering the nonmagnetic nature of $WTe_2$, we can exclude the possibility of magnetic-origin of this PHE. The PHE can be reproducibly detected in the as-grow $WTe_2$ single crystal. The PHE and anisotropic magnetoresistance (AMR) are well fitted by the formulas deduced from chiral anomaly, giving transport evidence of the topological nature in type-II Weyl semimetal.

## II. EXPERIMENTAL METHODS

Single crystals of $WTe_2$ were grown with Te flux. High-purity W and Te powder were placed in an alumina crucible sealed in an evacuated quartz tube. After the quartz ampoule was heated up to 1000°C, the ampoule was cooled down to 600°C at a rate of 2°C/h. The excess Te flux was removed by centrifugation at 600°C. Finally we can get large and layered crystals with clean surface. The crystal structure and phase purity were checked by single crystal X-ray diffraction (XRD) on a Rigaku-TTR3 X-ray diffractometer using Cu Kα radiation. Thin plate of single crystals $WTe_2$ were selected to perform the measurements. The longitudinal magnetoresistance (MR) and angle dependent planar Hall resistivity were taken on a Quantum Design PPMS. Standard four-probe technique was used to measure the longitudinal in-plane resistivity and Hall contacts were located on the transverse sides. The magnetic field was applied and rotated within the sample plane.

## III. RESULTS AND DISCUSSION

The obtained WTe$_2$ single crystal is of single phase and the naturally cleaved surface is the *ab* plane. In order to check the electronic properties, we take the temperature dependence of in-plane resistivity and longitudinal MR. As shown in Fig. 1(a), the temperature dependent in-plane resistivity exhibits a relatively large residual resistivity ratio, i.e., R(300 K)/R(2 K)=165, indicating the high quality of the sample. Figure 1(b) shows the longitudinal MR measured at various temperatures, with the magnetic field applied perpendicular to the current and sample plane. The MR reaches up to $10^5$% at 2 K and 14 T, without any sign of saturation, well reproducing the feature of extremely large MR [37]. For the curve taken at 2 K, clear Shubnikov–de Haas (SdH) oscillation signal can be detected when the applied magnetic field is larger than 5 T. After subtracting the non-oscillatory background from the MR data, the SdH oscillations are obtained, as shown in Fig. 1(c). Figure 1(d) shows the corresponding fast Fourier transformation (FFT) spectra at 2 K, in which several oscillation frequencies show up, i.e., 92 T ($\gamma$), 133 T ($\alpha$), 144 T ($\beta$), 165 T ($\delta$) and 277 T ($\alpha + \beta$), similar to previous reports [38, 39].

In order to measure the PHE, we take the experimental setup as theoretical work describes [31]. As illustrated in Fig. 2(a), the standard four-probe technique is used to measure the longitudinal in-plane resistivity and Hall contacts are located on the transverse sides. The magnetic field is applied and rotates within the ab plane. $\theta$ is defined as the angle of magnetic field with respect to electric current. The angle dependence of planar Hall resistivity ($\rho_{yx}$) and anisotropic MR ($\rho_{xx}$) are taken at various fields and temperatures. Figure 2(b) shows the angular dependence of $\rho_{yx}$ at 2 K and different fields, which is the average of the data taken under positive and negative fields. It is found that as the magnetic field increases, a non-zero Hall resistivity emerges. Note that, however, in actual experimental setup the field is not always strictly in the crystal plane, which causes a small angle ($\delta$) between the field-sweeping plane and crystal plane. Also, the Hall contacts are not perfectly symmetrical. Such two misalignments will induce a normal Hall resistivity and a small normal longitudinal resistivity in the measured Hall resistivity. The normal Hall resistivity has an angular dependence of $\sin\theta$, which would be cancelled by the averaged data of positive and negative fields. For a very small $\delta$, the included small normal longitudinal resistivity has an angular dependence of $\sin^2(\theta + \delta)$, considering that normal magnetoresistance has a quadratic field dependence. This term cannot be removed by data processing. However, noting that it is symmetrical for positive and negative fields, it is unlikely to account for the odd function type of measured $\rho_{yx}$. Another distinct feature of $\rho_{yx}$ is the angular period of π. Thus, we suggest that the measured $\rho_{yx}$ is an intrinsic property of WTe$_2$.

The PHE is a well-known phenomenon in ferromagnetic metals, which usually has a very small value and originates from the interplay of magnetic order and spin-orbit interactions. Considering the nonmagnetic nature of WTe$_2$, we exclude the possibility of any magnetic-origin of the PHE. We now demonstrate that the obtained giant PHE is due to the chiral anomaly, one nontrivial property arising from its topological nature. The PHE in TSMs is the appearance of a large transverse voltage when the in-plane magnetic field is not aligned with the current, which is formulated as [31]

$$\rho_{yx} = -\Delta\rho_{chiral}\sin\theta\cos\theta, \quad \rho_{xx} = \rho_\perp - \Delta\rho_{chiral}\cos^2\theta, \quad (1)$$

where $\Delta\rho_{chiral} = \rho_\perp - \rho_\parallel$ is the chiral anomaly induced resistivity anisotropy, $\rho_\perp$ and $\rho_\parallel$ are the resistivity corresponding to the current flow perpendicular to and along the direction of magnetic field, respectively. In order to reveal the underlying information, we use the chiral anomaly model to fit the PHE data in Fig. 2(b). With or without a normal longitudinal resistivity term, both methods are tried and give nearly identical fitting results, suggesting that the misalignments mentioned above are negligible in our measurements. The solid red curves in Fig. 2(b) represent the fittings to the formula of $\rho_{yx}$. The experimental data and fitted data agree with each other very well, which means that the measured PHE in WTe$_2$ is originating from the chiral anomaly.

In order to check the temperature dependence of PHE, we carry out the PHE measurements at various temperatures under 14 T. Generally speaking, the chiral anomaly in TSMs is insensitive to temperature, which may be persistent to very high temperature. As shown in Fig. 2(c), the PHE can be seen up to 200 K, which is consistent with the behavior of chiral anomaly induced NMR in TSMs [14, 25]. Figure 2(d) presents the temperature dependence of $\Delta\rho_{chiral}$. The gradual decrease of $\Delta\rho_{chiral}$ reflects the influence of thermal fluctuation.

According to the theory of chiral anomaly induced PHE, the AMR $\rho_{xx}$ is particular in TSMs, as described in Eq. (1). We measure the angular dependence of $\rho_{xx}$ at 2 K under various fields. As shown in Fig. 3(a), as the field increases, a pronounced longitudinal resistivity is observed, with a period of π. Solid red curves represent the fittings using Eq. (1). Here, we neglect the small normal longitudinal resistivity that is due to the possible small angle $\delta$ between the field-sweeping plane and the sample plane. Again, the experimental data can be fitted well, which means that the experimental misalignments in our measurements are small and the resistivity anisotropy induced by chiral anomaly is prominent. The field dependence of fitting parameters at 2 K are shown in Fig. 3(b). The obtained $\Delta\rho_{chiral}$ and $\rho_\perp$ both increase with field, and $\Delta\rho_{chiral}$ contributes most of $\rho_\perp$ ($\rho_\perp = \Delta\rho_{chiral} + \rho_\parallel$), suggesting that $\rho_\parallel$ is nearly zero at 2 K.

Figure 3(c) shows the angular dependence of $\rho_{xx}$ at 14 T for various temperatures, and the fitting parameters as the function of temperature are summarized in Fig. 3(d). As the temperature increases, the resistivity anisotropy becomes weaker, consistent with the temperature dependence of $\Delta\rho_{chiral}$ in Fig. 2(d). This means that the AMR and PHE in topological semimetals share the same mechanism, i.e., chiral anomaly, confirming the topological nature of WTe$_2$. Moreover, the difference of $\rho_\perp$ and $\Delta\rho_{chiral}$ (i.e., $\rho_\parallel$) gradually increases with temperature, until approaches the value of $\rho_\perp$ at high temperature.

## IV. CONCLUSIONS

In conclusion, we report the PHE in a type-II Weyl semimetal for the first time, WTe$_2$. The measured AMR and PHE can be well fitted by the formulas deduced from chiral anomaly. Our experiment demonstrates that the chiral anomaly is robust existing in WTe$_2$. Our work provides new evidence for the topological semimetal state of WTe$_2$, and confirms that the PHE is a universal and powerful tool to determine the topological nature of TSMs.

**ACKNOWLEDGMENTS**

This work was supported by the National Key R&D Program of China (Grant Nos. 2016YFA0300404 and 2017YFA0403600), the National Natural Science Foundation of China (Grant Nos. 51603207, U1532267 and 11674327), and the Natural Science Foundation of Anhui Province (Grant No. 1708085MA08). Y.J.W. and J.X.G. contributed equally to this work.

[1] X. Wan, A. M. Turner, A. Vishwanath, S.Y. Savrasov, Topological semimetal and Fermi-arc surface states in the electronic structure of pyrochlore iridates, Physical Review B, 83 (2011) 205101.

[2] A. A. Burkov, Topological semimetals, Nat Mater, 15 (2016) 1145-1148.

[3] S. M. Young, S. Zaheer, J. C. Teo, C. L. Kane, E. J. Mele, A. M. Rappe, Dirac semimetal in three dimensions, Phys Rev Lett, 108 (2012) 140405.

[4] Z. Wang, Y. Sun, X. Q. Chen, C. Franchini, G. Xu, H. Weng, X. Dai, Z. Fang, Dirac semimetal and topological phase transitions in $A_3Bi$ (A=Na, K, Rb), Physical Review B, 85 (2012) 195320.

[5] Z. Wang, H. Weng, Q. Wu, X. Dai, Z. Fang, Three-dimensional Dirac semimetal and quantum transport in $Cd_3As_2$, Physical Review B, 88 (2013) 125427.

[6] S. Borisenko, Q. Gibson, D. Evtushinsky, V. Zabolotnyy, B. Buchner, R. J. Cava, Experimental realization of a three-dimensional Dirac semimetal, Physical review letters, 113 (2014) 027603.

[7] S. M. Huang, S. Y. Xu, I. Belopolski, C. C. Lee, G. Chang, B. Wang, N. Alidoust, G. Bian, M. Neupane, C. Zhang, S. Jia, A. Bansil, H. Lin, M. Z. Hasan, A Weyl Fermion semimetal with surface Fermi arcs in the transition metal monopnictide TaAs class, Nature communications, 6 (2015) 7373.

[8] H. Weng, C. Fang, Z. Fang, B. A. Bernevig, X. Dai, Weyl Semimetal Phase in Noncentrosymmetric Transition-Metal Monophosphides, Physical Review X, 5 (2015) 011029.

[9] S. Y. Xu, I. Belopolski, N. Alidoust, M. Neupane, G. Bian, C. Zhang, R. Sankar, G. Chang, Z. Yuan, C. C. Lee, S. M. Huang, H. Zheng, J. Ma, D. S. Sanchez, B. Wang, A. Bansil, F. Chou, P. P. Shibayev, H. Lin, S. Jia, M. Z. Hasan, Discovery of a Weyl fermion semimetal and topological Fermi arcs, Science, 349 (2015) 613.

[10] B. Q. Lv, H. M. Weng, B. B. Fu, X. P. Wang, H. Miao, J. Ma, P. Richard, X. C. Huang, L. X. Zhao, G. F. Chen, Z. Fang, X. Dai, T. Qian, H. Ding, Experimental Discovery of Weyl Semimetal TaAs, Physical Review X, 5 (2015) 031013.

[11] Y. Wu, D. Mou, N. H. Jo, K. Sun, L. Huang, S. L. Bud'ko, P. C. Canfield, A. Kaminski, Observation of Fermi arcs in the type-II Weyl semimetal candidate $WTe_2$, Physical Review B, 94 (2016) 121113.

[12] A. A. Soluyanov, D. Gresch, Z. Wang, Q. Wu, M. Troyer, X. Dai, B. A. Bernevig, Type-II Weyl semimetals, Nature, 527 (2015) 495.

[13] L. P. He, X. C. Hong, J. K. Dong, J. Pan, Z. Zhang, J. Zhang, S.Y. Li, Quantum transport evidence for the three-dimensional Dirac semimetal phase in $Cd_3As_2$, Phys Rev Lett, 113 (2014) 246402.

[14] X. Huang, L. Zhao, Y. Long, P. Wang, D. Chen, Z. Yang, H. Liang, M. Xue, H. Weng, Z. Fang, X. Dai, G. Chen, Observation of the Chiral-Anomaly-Induced Negative Magnetoresistance in 3D Weyl Semimetal TaAs, Physical Review X, 5 (2015) 031023.

[15] A. A. Burkov, Chiral anomaly and diffusive magnetotransport in Weyl metals, Phys Rev Lett 113 (2014) 247203.

[16] A. A. Burkov, Negative longitudinal magnetoresistance in Dirac and Weyl metals, Physical Review B, 91 (2015) 245157.

[17] H. Li, H. He, H. Z. Lu, H. Zhang, H. Liu, R. Ma, Z. Fan, S. Q. Shen, J. Wang, Negative magnetoresistance in Dirac semimetal $Cd_3As_2$, Nature communications, 7 (2016) 10301.

[18] K. Fukushima, D. E. Kharzeev, H. J. Warringa, Chiral magnetic effect, Physical Review D, 78 (2008) 074033.

[19] K. Y. Yang, Y. M. Lu, Y. Ran, Quantum Hall effects in a Weyl semimetal: Possible application in pyrochlore iridates, Physical Review B, 84 (2011) 075129.

[20] G. Xu, H. Weng, Z. Wang, X. Dai, Z. Fang, Chern Semimetal and the Quantized Anomalous Hall Effect in $HgCr_2Se_4$, Physical review letters, 107 (2011) 186806.

[21] S. A. Parameswaran, T. Grover, D. A. Abanin, D. A. Pesin, A. Vishwanath, Probing the Chiral Anomaly

with Nonlocal Transport in Three-Dimensional Topological Semimetals, Physical Review X, 4 (2014) 031035.

[22] C. Zhang, E. Zhang, W. Wang, Y. Liu, Z. G. Chen, S. Lu, S. Liang, J. Cao, X. Yuan, L. Tang, Q. Li, C. Zhou, T. Gu, Y. Wu, J. Zou, F. Xiu, Room-temperature chiral charge pumping in Dirac semimetals, Nature communications, 8 (2017) 13741.

[23] S. K. K. Jun Xiong, Tian Liang, Jason W. Krizan, Max Hirschberger, Wudi Wang, R. J. Cava, N. P. Ong, Evidence for the chiral anomaly in the Dirac semimetal Na3Bi, Science, 350 (2015) 413.

[24] Q. Li, D. E. Kharzeev, C. Zhang, Y. Huang, I. Pletikosić, A.V. Fedorov, R.D. Zhong, J.A. Schneeloch, G.D. Gu, T. Valla, Chiral magnetic effect in ZrTe5, Nature Physics, 12 (2016) 550-554.

[25] Y. Wang, E. Liu, H. Liu, Y. Pan, L. Zhang, J. Zeng, Y. Fu, M. Wang, K. Xu, Z. Huang, Z. Wang, H.Z. Lu, D. Xing, B. Wang, X. Wan, F. Miao, Gate-tunable negative longitudinal magnetoresistance in the predicted type-II Weyl semimetal WTe2, Nature communications, 7 (2016) 13142.

[26] M. Hirschberger, S. Kushwaha, Z. Wang, Q. Gibson, S. Liang, C.A. Belvin, B.A. Bernevig, R.J. Cava, N.P. Ong, The chiral anomaly and thermopower of Weyl fermions in the half-Heusler GdPtBi, Nat Mater, 15 (2016) 1161-1165.

[27] Y. Y. Lv, X. Li, B. B. Zhang, W. Y. Deng, S. H. Yao, Y. B. Chen, J. Zhou, S. T. Zhang, M. H. Lu, L. Zhang, M. Tian, L. Sheng, Y. F. Chen, Experimental Observation of Anisotropic Adler-Bell-Jackiw Anomaly in Type-II Weyl Semimetal WTe1.98 Crystals at the Quasiclassical Regime, Physical review letters, 118 (2017) 096603.

[28] L. Ritchie, G. Xiao, Y. Ji, T.Y. Chen, C.L. Chien, M. Zhang, J. Chen, Z. Liu, G. Wu, X.X. Zhang, Magnetic, structural, and transport properties of the Heusler alloys Co2MnSi and NiMnSb, Physical Review B, 68 (2003) 104430.

[29] J. P. Ulmet, L. Bachère, S. Askenazy, J. C. Ousset, Negative magnetoresistance in some dimethyltrimethylene-tetraselenafulvalenium salts: A signature of weak-localization effects, Physical Review B, 38 (1988) 7782.

[30] S. Nandy, G. Sharma, A. Taraphder, S. Tewari, Chiral Anomaly as the Origin of the Planar Hall Effect in Weyl Semimetals, Physical review letters, 119 (2017) 176804.

[31] A. A. Burkov, Giant planar Hall effect in topological metals, Physical Review B, 96 (2017) 041110.

[32] Hui Li, Huanwen Wang, Hongtao He, Jiannong Wang, Shun-Qing Shen, Giant Anisotropic Magnetoresistance and Planar Hall Effect in Dirac Semimetal Cd3As2, arXiv:1711.03671.

[33] Min Wu, Guolin Zheng, Weiwei Chu, Wenshuai Gao, Hongwei Zhang, Jianwei Lu, Yuyan Han, Jiyong Yang, Haifeng Du, Wei Ning, Yuheng, Z.a.M. Tian, Probing the Chiral Anomaly by Planar Hall Effect in Three-dimensional Dirac Semimetal Cd3As2 Nanoplates, arXiv:1710.01855.

[34] Nitesh Kumar, Claudia Felser, Chandra Shekhar, Planar Hall effect in Weyl semimetal GdPtBi, arXiv:1711.04133.

[35] A. M. Nazmul, H. T. Lin, S.N. Tran, S. Ohya, M. Tanaka, Planar Hall effect and uniaxial in-plane magnetic anisotropy in Mn δ-doped GaAs/p-AlGaAs heterostructures Physical Review B, 77 (2008) 155203.

[36] F. Y. Bruno, A. Tamai, Q.S. Wu, I. Cucchi, C. Barreteau, A. de la Torre, S. McKeown Walker, S. Riccò, Z. Wang, T.K. Kim, M. Hoesch, M. Shi, N.C. Plumb, E. Giannini, A.A. Soluyanov, F. Baumberger, Observation of large topologically trivial Fermi arcs in the candidate type-II Weyl semimetal WTe2, Physical Review B, 94 (2016) 121112.

[37] M. N. Ali, J. Xiong, S. Flynn, J. Tao, Q. D. Gibson, L. M. Schoop, T. Liang, N. Haldolaarachchige, M. Hirschberger, N. P. Ong, R. J. Cava, Large, non-saturating magnetoresistance in WTe2, Nature, 514 (2014)

205.

[38] P. L. Cai, J. Hu, L. P. He, J. Pan, X. C. Hong, Z. Zhang, J. Zhang, J. Wei, Z. Q. Mao, S. Y. Li, Drastic Pressure Effect on the Extremely Large Magnetoresistance in WTe2: Quantum Oscillation Study, Physical review letters, 115 (2015) 057202.

[39] Z. Zhu, X. Lin, J. Liu, B. Fauqué, Q. Tao, C. Yang, Y. Shi, K. Behnia, Quantum Oscillations, Thermoelectric Coefficients, and the Fermi Surface of Semimetallic WTe2, Physical review letters, 114 (2015) 176601.

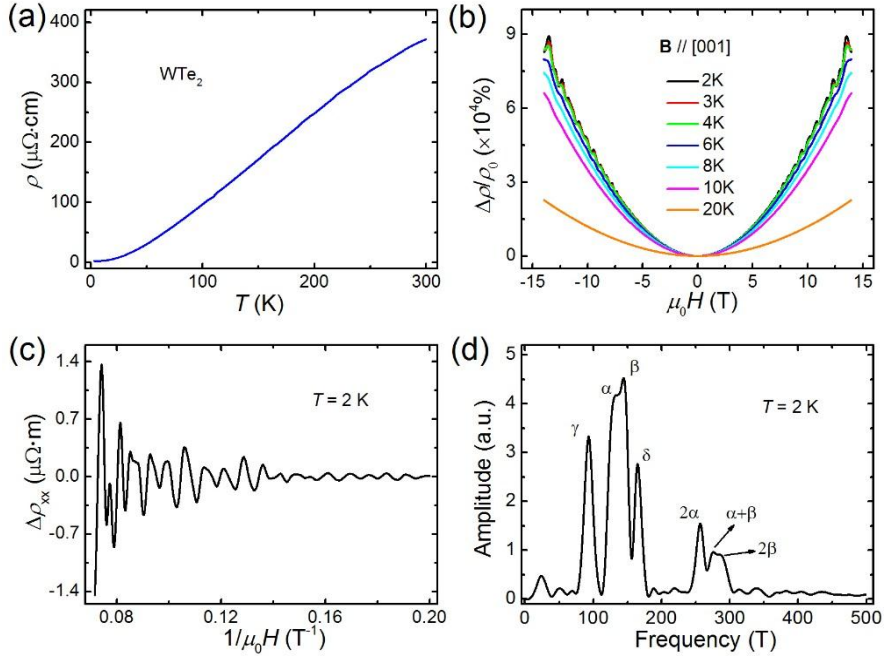

Fig. 1. (a) Temperature dependence of in-plane resistivity. (b) Magnetoresistance measured at different temperatures, with the magnetic field applied perpendicular to the current and sample plane. (c) SdH oscillations at 2 K, obtained by subtracting the non-oscillatory background from the MR data. (d) FFT spectra of the SdH oscillations in (c).

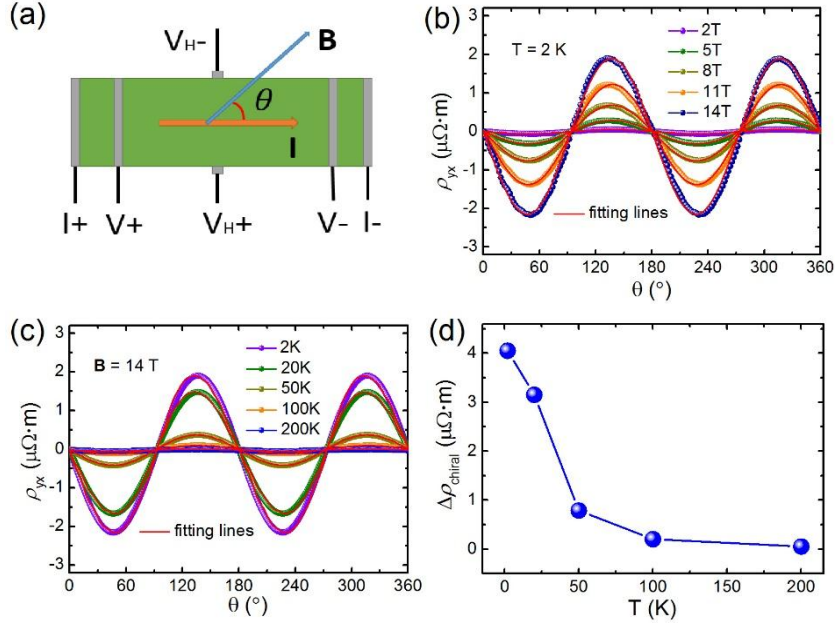

Fig. 2. (a) Schematic illustration for the PHE measurement geometry. Standard four-probe technique is used to measure the longitudinal in-plane resistivity and Hall contacts are located on the transverse sides. The magnetic field is applied and rotates within the sample plane. $\theta$ is defined as the angle of magnetic field with respect to electric current. (b) Angle dependence of $\rho_{yx}$ taken at 2 K for various fields. (c) Angle dependence of $\rho_{yx}$ taken at 14 T for various temperatures. Solid red curves represent the fittings to chiral anomaly formula. (d) Temperature dependence of $\Delta\rho_{chiral}$ at 14 T obtained from the fitting results in (c).

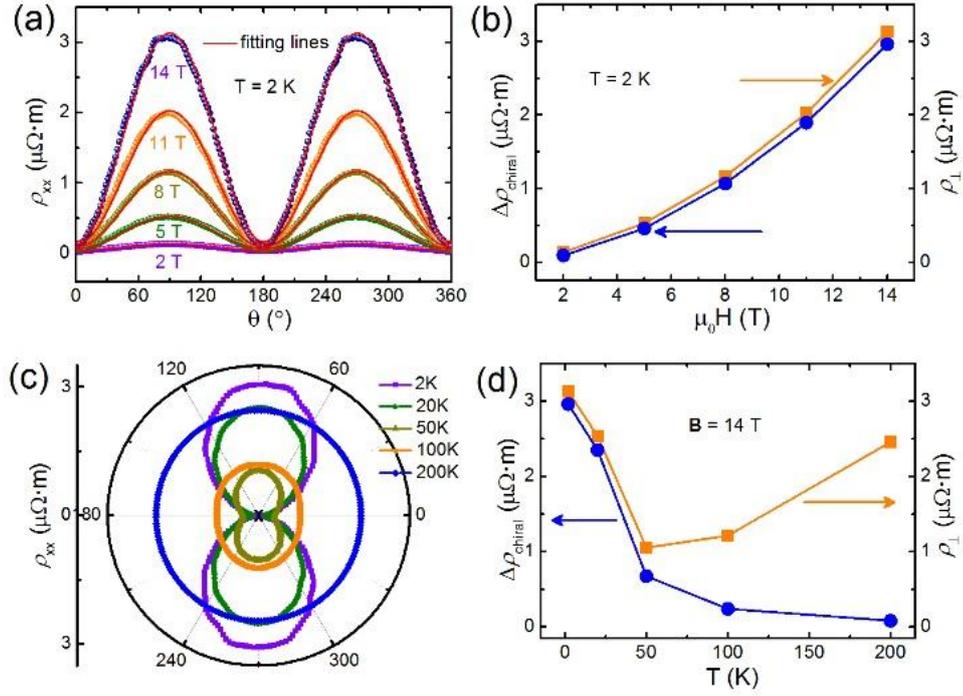

Fig. 3. (a) Angle dependence of $\rho_{xx}$ taken at 2 K for various fields. Solid red curves represent the fittings to chiral anomaly formula. (b) Field dependence of $\Delta\rho_{chiral}$ and $\rho_\perp$ at 2 K obtained from the fitting results in (a). (c) Angle dependence of $\rho_{xx}$ taken at 14 T for various temperatures. (d) Temperature dependence of $\Delta\rho_{chiral}$ and $\rho_\perp$ at 14 T obtained from the fitting results in (c).